\input psfig
\documentstyle[12pt,aaspp4]{article}
\begin{document}
\title{Screening in Thermonuclear Reaction Rates in the Sun}
\author{Andrei V. Gruzinov and John N. Bahcall}

\affil{Institute for Advanced Study, School of Natural Sciences,
Princeton, NJ 08540}

\begin{abstract}
We evaluate the effect of  electrostatic screening by ions
and electrons 
on low-Z thermonuclear reactions in the sun. 
We use a mean field formalism and calculate
the electron density  of the screening cloud 
using the appropriate density matrix equation
of quantum statistical mechanics.
Because of well understood physical effects that are 
included for the first time in our treatment,
the calculated enhancement of reaction rates
does not agree with the frequently used interpolation 
formulae. Our result does agree, within small uncertainties, with 
Salpeter's weak screening formula. If weak screening is used 
instead of the commonly employed screening prescription of 
Graboske et al., the predicted $^8$B neutrino flux is increased 
by 7\%  and the predicted chlorine rate is increased by 0.4 SNU.

\end{abstract}

\section{Introduction}

In recent years, an increasing amount of attention has been 
devoted to calculating more accurately the effects on the rates
of solar fusion reactions of electrostatic screening in the solar
plasma (Carraro, Sch\"afer, \& Koonin 1988; Johnston, Kolbe, Koonin,
\& Langanke 1992; Bahcall \& Pinsonneault 1992; Shoppa, Koonin,
Langanke, \& Seki 1993; Dzitko, Turck-Chi\`eze, Delbourgo-Salvador, \&
 Lagrange 1995; Ricci, Degl'Innocenti, \& Fiorentini 1995; Gruzinov \&
 Bahcall 1997; Brown \& Sawyer 1997; Br\"uggen \& Gough 1997). 
All of these discussions take as their starting point the classical
 analysis by Salpeter (1954). The primary reason for making 
more precise calculations
 is that nuclear fusion reactions produce the solar neutrino fluxes (see http://www.sns.ias.edu/~jnb ). 
The neutrino fluxes are being observed with a number of large 
new detectors that are expected to yield flux measurements of 
high accuracy (of the order of a few percent or better, see Bahcall
 et al. 1995, and, for more details, 
Totsuka 1996; McDonald 1994; Arpesella et al. 1992).

In this paper, we calculate for the first time
the electron density in the vicinity of the fusing nuclei using the 
partial differential equation for the density matrix that is derived
in quantum statistical mechanics.
In previous treatments of screening that attempted to go beyond the
linear regime, the electron density near the nucleus was either taken
to be--without quantitative justification--the unperturbed
 value, $n_e(\infty)$ (Mitler 1977; 
Dzitko, Turck-Chi\`eze, Delbourgo-Salvador, \& Lagrange 1995) 
or left as a free parameter
(Ricci et al. 1995) or the electrons were assumed to be completely
degenerate (Graboske, DeWitt, 
Grossman, \& Cooper 1973). 
We calculate screening 
corrections in a mean field approximation; 
we numerically solve the nonlinear Poisson-Boltzmann equation for a
mixture of electrons and ions.
The electron density distribution calculated from the density matrix
equation is included self-consistently and iteratively in the mean
field equation.

Our results
represent both an improvement on, and a simplification of, the
description of nuclear fusion used in many solar evolution codes.

For simple physical reasons, our results differ from the 
interpolation formulae that are currently used to describe reaction 
rates in the Sun (Salpeter \& Van Horn 1969; Graboske et al.
1973), and the numerical calculations of 
Dzitko, Turck-Chi\`eze, Delbourgo-Salvador, \& Lagrange (1995). 

Interpolation formulae describe a transition between Salpeter's 
weak screening, which is due to both electrons and ions, and strong 
screening, for which only ions are effective. At the high densities
relevant for strong screening,  electrons are fully 
degenerate. The solar core, however, is only weakly 
degenerate, and the effects of degeneracy are already included into 
the Debye radius, $R_D$ (which is increased $2$\% by electron
degeneracy, cf. Eq.~(\ref{defnzeta}) of the present paper or  
Eq.~(25) of Salpeter 1954). 
Therefore, the interpolation formulae 
in use underestimate the electron contribution to screening and give 
reaction rates lower than ours. 

The numerical procedures of Dzitko, Turck-Chi\`eze, Delbourgo-Salvador, 
\& Lagrange (1995) and Mitler (1977) predict
 reaction rates that are too slow 
for heavy ions because  they  assumed 
that the 
electron charge density near a screened nucleus
is  the unperturbed value, 
$en_e(\infty)$.
This assumption seriously underestimates the charge density near heavy ions.
For example, it is  known that 
a screened beryllium nucleus under solar interior
conditions has charge density near the nucleus $\approx -3.85en_e(\infty)$ 
(Gruzinov \& Bahcall 1997, Brown \& Sawyer 1997; all   quantum
mechanical calculations
give similar results, see Bahcall 1962 and Iben, Kalata, \&
Schwartz 1967).

This paper is organized as follows.
In \S 2 we review the basic concepts and  in \S 3 we relate the 
electrostatic energy to the screening enhancement using the 
free energy. We describe the calculations  in \S 4 and summarize 
the numerical results in \S 5. 
In \S 6 we summarize our main results and present the conclusions
regarding solar neutrino fluxes.
The Appendix evaluates a quantum correction 
to the kinetic energy of thermal electrons in the electrostatic 
field of a screened nucleus.

\section{Enhancement of Fusion Rates}

The solar core plasma is  dense enough that it 
 noticeably enhances fusion 
rates as compared to the rates in a rarefied plasma of the same 
temperature. As explained by Salpeter (1954), the rate of a fusion of
two nuclei of charges $Z_1$ and $Z_2$ is increased by a factor
\begin{equation}
\label{deff}
f=\exp \Lambda,
\end{equation}
where
\begin{equation}
\label{deflambda}
\Lambda =Z_1Z_2{e^2\over TR_D}.
\end{equation}
Here $R_D$ is the Debye radius,
\begin{equation}
\label{defndebye}
{1\over R_D^2}=4\pi\beta ne^2\zeta^2 ,
\end{equation}
with
\begin{equation}
\label{defnzeta}
\zeta = \left\{\Sigma_i X_i {Z^2_i\over A_i} + \left({f^\prime\over
f}\right) \Sigma_i X_i {Z_i\over A_i}\right\}^{1/2} .
\end{equation}

Here $\beta = 1/T$, $n$ is the baryon density, $X_i, Z_i$, and $A_i$
are, respectively, the mass fraction, the nuclear charge, and the
atomic weight of ions of type $i$.  The quantity $f^\prime/f \simeq
0.92$ accounts for electron degeneracy.  Equation (\ref{defnzeta}) is
the same as Eq.~(25) of Salpeter (1954).  In what follows, we will
make use of a simplified expression for $\zeta$,
\begin{equation}
\label{zetasimple}
\zeta_{\rm simple} \cong \left\{\left(1 - Y/2\right) f^\prime/f +
1\right\}^{1/2} ,
\end{equation}
in which the plasma is assumed to consist only of hydrogen
and helium ($Y$ is the helium
abundance by mass).  The approximation of considering only a hydrogen
and helium plasma rather than the full solar composition (cf. Grevesse
 \& Noels 1993) causes an error of less than 0.5\% in computing solar
fusion rates.  This error is completely unimportant for our
purpose of estimating the ratio of the total screening to the weak
screening value given by Eqs. (\ref{deff})--(\ref{defnzeta}).

The enhancement of fusion rates due to screening depends only very
weakly upon location in the solar interior (cf. Ricci et al. 1995), 
because the primary effect
of screening is proportional to $\rho/T^3$ 
(cf. Eq.~\ref{deff}--\ref{defnzeta}),
which is approximately constant in the solar interior (cf. Eq.~41 of
Bahcall et al. 1982). 
The plasma parameters 
at a characteristic radius in the solar interior, 
 $R/R_{\odot }=0.06$, are (Bahcall \& Pinsonneault 1995) 
$R_D = 0.46$ and $T = 47$ in atomic units ($m_e=\hbar =e=1$). 
In units that are more common in astronomical discussions, 
the temperature, $T_6$, in millions of degrees is
$T_6=0.32T=15$ and the Debye radius in cm is $R_D =
0.46 \times 5.3\times 10^{-9} = 2\times 10^{-9}$ cm.

Consider an important example: $Z_1Z_2=4$ for the solar 
fusion reactions $^3{\rm He}(^4{\rm He},\gamma)^7{\rm Be}$ 
and $^7{\rm Be}({\rm p},\gamma)^8{\rm B}$. 
Eq. (2) yields $\Lambda =0.19$. According to Eq. (1), the calculated
rates of these fusion reactions are then 21\% faster than they 
would be if screening were neglected.

Eq. (1) is only valid to first order in $\Lambda$. Nonlinearities 
in the electrostatic screening interactions 
might naively be expected to produce
corrections 
$\sim \Lambda ^2$, i.e., of order 4\% for a $Z_1Z_2=4$ reaction. 
In the following sections, we calculate corrections to the Salpeter
weak screening formula, Eq.~(\ref{deff}), and find that the 
numerical corrections
are always significantly smaller than $\Lambda ^2$. 

\section{Enhancement Factors and Free Energy}

Salpeter's formula (Eqs. (1),(2)) can be derived as follows. The screened 
potential near the nucleus $Z_1$ in the Debye-H\"uckel approximation is
\begin{equation}
\label{Zshift}
{Z_1\over r}e^{-r/R_D}\approx {Z_1\over r}-{Z_1\over R_D}.
\end{equation}
The potential shift $Z_1/R_D$ increases the probability that the charge $Z_2$ 
comes close to $Z_1$ by the Boltzmann factor $e^{\Lambda }$, $\Lambda =\beta 
Z_1Z_2/R_D$. 

Unfortunately, this clear derivation can not be used if we go beyond the 
Debye-H\"uckel approximation and include nonlinear screening effects. Given a 
numerically calculated potential around the charge $Z_1$, $\phi _1(r)$ , we 
can not assume that the enhancement factor is equal to $e^{\Lambda }$ with 
$\Lambda =\beta Z_2\times (Z_1/r-\phi_1(r))|_{r=0}$. This is already obvious 
from the asymmetry of this expression under the 1-2 permutation; $\phi _1$ is 
not just proportional to $Z_1$ for nonlinear screening. 

In the more general case considered here, 
the enhancement of fusion rates due to screening can be
calculated in terms of an expression involving the free energy of 
a screened charge $Z$, $F(Z)$. In terms of free energy, the 
enhancement factor is simply (DeWitt, Graboske, \& Cooper 1973) $e^{\Lambda }$ 
with 
\begin{equation}
\label{lambdasym}
\Lambda =-\beta F(Z_1+Z_2)+\beta F(Z_1)+\beta F(Z_2),
\end{equation}
which is a manifestly symmetrical expression. 
Equation(\ref{lambdasym}) expresses the thermodynamic relation that at
constant temperature (which is relevant when considering solar fusion
reactions) $\delta F = -\delta W$, where $\delta W$ is the work done 
by the plasma on the fusing ions.
The extra work performed by the plasma
due to screening is positive, pushing the fusing ions
closer together. For a given relative kinetic energy when the ions
fuse, the initial kinetic energy is lower by $\delta W$ 
than in the absence of
screening. 
Therefore, the probability of the fusing configuration is increased,
i.e., the reaction rates are faster, by a factor $\exp(\beta \delta W/T) =
\exp(-\beta \delta F) = \exp(\Lambda)$.

The free energy can be calculated in 
terms of electrostatic energy using the thermodynamic formula 
\begin{equation}
\label{FintermsofU}
\beta F=\int_0^{\beta }d\beta ' U .
\end{equation}
The lower limit in the integral in Eq. (\ref{FintermsofU}) is chosen
so that at high-temperature (small $\beta$) $F$ goes to zero as
$\beta^{1/2}$, as implied by Debye theory (see discussion below).
The total electrostatic energy including the self-energy is 
\begin{equation}
\label{Ufull}
U_{\rm tot} = {1\over 2}\int d^3r\phi (r) \rho (r).
\end{equation}
The self-energy of the charges cancels out in performing the
difference indicated by \hbox{Eq. (\ref{lambdasym})}. 
The fusing nuclei are  well
separated whenever  screening is relevant;  their combined self-energies are
the same in the fusing state as the sum of the self-energies in the
initial (infinitely separated) state. 
Most of the acceleration of the fusing nuclei occurs at distances
larger than $0.1 R_{D}$, which is four orders of magnitude larger than     
nuclear radii.
Therefore, the relevant self-energy for the calculation of enhancement
factors due to screening does not include the self-energies and is
\begin{equation}
\label{Ucorrected}
U = {Z\over 2}\delta \phi (0)+{1\over 2}\int d^3r\phi (r) \delta \rho (r),
\end{equation}
where $\delta \phi =\phi-Z/r$ and $\delta \rho =\rho -Z\delta ({\bf r})$.

In the Debye-H\"uckel approximation these expressions reproduce 
Salpeter's formula (Br\"uggen \& Gough  1997). In the 
Debye-H\"uckel approximation, 
$\phi =(Z/r)e^{-r/R_D}$, $4\pi \delta \rho =-\phi /R_D^2$, 
$\delta \phi (0)=-Z/R_D$, and Eq. (\ref{Ucorrected}) 
gives $U=-(3/4)Z^2/R_D$. Since $R_D\sim \beta ^{-1/2}$, 
Eq. (\ref{FintermsofU}) 
gives $\beta F=-(1/2)\beta Z^2/R_D$. Then Eq. (\ref{lambdasym}) 
gives Eq. (\ref{deflambda}).

\section{Calculations}
\label{seccalculations}

In the mean field approximation, electrostatic screening of a 
charge $Z$ is described by the Poisson-Boltzmann equation
\begin{equation}
\label{eqnPB}
\nabla ^2\phi = 
4\pi n\{ (1-{Y\over 2})e^{\beta \phi}-(1-Y)e^{-\beta \phi}-
{Y\over 2}e^{-2\beta \phi}\} ,
\end{equation}
where the terms on the right hand side represent, respectively, 
screening by electrons, protons and alphas. The boundary 
condition is $\phi \rightarrow Z/r$ for $r \rightarrow 0$. 
In the non-linear regime, 
one cannot solve the Poisson-Boltzmann equation as written. 
Classical electrons recombine,  which 
corresponds formally to the  divergence of the classical Boltzmann 
factor $e^{\beta \phi }$ near the nucleus. 
This problem does not arise in previous solutions of the
Poisson-Boltzmann equation which were carried out in the linear regime 
corresponding to weak
screening. 

Quantum statistical
mechanics 
must be used for calculating terms beyond the weak screening
approximation. Fortunately, electron degeneracy makes only a small
correction, less than $2$\% in the Debye radius, see
Eq.~\ref{defnzeta}, and therefore less than $1$\% for all cases in the
reaction rates. Hence,  
a distinguishable 
particles approximation can be employed\footnote{Brown and Sawyer
(1997) calculated electron densities using both Fermi-Dirac and
Maxwell-Boltzmann statistics.  Degeneracy effects on the value of the
central electron density were of order 10\% for Z = 6.  We shall
show in the course of this paper that changing the central electron
density by almost an order of magnitude does not significantly change
the rate of nuclear fusion reactions. Therefore, the small fractional
change in the central electron density due to using different
statistics is not important for our purposes. }. 
We use a numerical 
code that solves the density matrix equation for the density of
electrons near the nucleus. The code, which was developed following
the discussion of Feynman (1990), is described in Gruzinov \& Bahcall
(1997).

The average electron density can be 
calculated by solving the density matrix equation (e.g. Feynman 1990)
\begin{equation}
\label{densitymatrixeq}
\partial _\beta \rho =\{ {1\over 2}\nabla ^2 + \phi(r) \} \rho, 
\end{equation}
with the initial condition
\begin{equation}
\label{boundarycondition}
\rho (r,\beta =0)=\delta ^{(3)}(r).
\end{equation}
Since Eq.~(\ref{densitymatrixeq}) appropriately describes the quantum
statistical mechanical effects, the solution for the density matrix
converges everywhere despite the divergence of the classical potential
at $r = 0$.
Another great advantages of the density matrix formulation is that
the character of the states in the plasma does not have to be
specified and therefore difficult questions concerning the existence
or non-existence of bound states are finessed.
The enhancement of the electron density to be used in the Poisson-Boltzmann equation instead of the Boltzmann factor $e^{\beta \phi}$ is
the solution of Eq.~(\ref{densitymatrixeq}) for the nuclear charge of 
Z divided by the solution for $Z = 0$.  The solution for the $Z = 0$
case can be obtained analytically and is
 $\rho _0(\beta)=(2\pi \beta 
)^{-3/2}$.

As described in Gruzinov \& Bahcall (1997),
the diffusion with multiplication problem, Eq. (\ref{densitymatrixeq}), 
can be solved easily by 
direct three-dimensional numerical simulations for solar conditions, because 
the 
inverse temperature $\beta$ is small ($\sim 0.02$), and the diffusive 
trajectory 
stays close to the origin. 
The mesh size and the regularization procedure were the same as in our
previous work.

Numerically, we  start with an initial guess that 
 $\phi (r)=(Z/r)e^{-r/R_D}$ everywhere and then calculate the 
electron density using  Eq.~(\ref{densitymatrixeq}) for all
$r < 0.4$.
The particular value of $ r = 0.4$ ($\sim R_D$) 
is not important. For all $r\gtrsim 0.2$, our 
density matrix code simply reproduces the Boltzmann distribution
factor $n(r) = n(\infty) e^{\beta \phi}$. 
We use the calculated electron density 
at $r<0.4$ to solve Eq. (\ref{eqnPB}) numerically for all $r$. We then
obtain a 
new potential $\phi (r)$. We use this potential to calculate the 
electron density at $r<0.4$ using Eq.~(\ref{densitymatrixeq})
and repeat
 the procedure.
The procedure converges quickly, after one to three iterations. 

The 
electrostatic energy was calculated from Eq. (\ref{Ucorrected}). The 
calculation was repeated at higher temperatures for the purpose of 
estimating the free energy using Eq. (\ref{FintermsofU}). 

Quantum statistical mechanics implies the existence of an effect that
we believe has not been previously considered in the context of fusion
reaction rates. The kinetic energy of electrons in the electrostatic
field of the nucleus is no longer ${3\over 2}T$ per electron. Indeed,
the kinetic energy of electrons is increased. In the low-temperature
limit this effect is the familiar zero-point oscillations. In the high
temperature limit the effect is more subtle, but it can be calculated
analytically. In the Appendix,
 we calculate the quantum statistical mechanics corrections to  the 
electron kinetic energy and the resulting correction to the free energy.     

\section{Numerical Results}
\label{secnumericalresults}

Table 1 gives the numerical results for: (i) corrections to 
the Debye-H\"uckel electrostatic energy, (ii) corrections to the 
free energy due to the changed electrostatic energy, (iii) corrections
to the kinetic energy of electrons, (iv) corrections to the free 
energy due to the changed kinetic energy of electrons, and (v) the total 
correction to free energy.

 Figure 1 explains the sources of different corrections to the 
Debye-H\"uckel approximation of screening. (i) At 
large distances (small $\phi$), the plasma response is 
suppressed due to helium ions. To see this, expand 
Eq. (\ref{eqnPB}) up to the second order 
in $\phi$: $\nabla ^2\phi=\phi (1-w\beta  \phi)/R_D^2$, 
where $w=3Y/(8-2Y)$.  (ii) At small 
distances, the plasma response is suppressed due to 
the fuzziness of quantum electrons, which is expressed by the density
matrix Eq.~(\ref{densitymatrixeq}).  (iii) At 
intermediate radii, the plasma response can be 
enhanced just because $e^{\beta \phi}>\beta \phi$. 

The second column of Table~2 shows the corrections, $-\delta \Lambda$, 
to reaction rates calculated in 
this paper relative to Salpeter's weak screening rates. For 
example, the correction to the rate of the 
$^7{\rm Be}({\rm p},\gamma)^8{\rm B}$ reaction is 
$\delta \Lambda =-\beta \delta F(5)+\beta \delta F(4)+\beta \delta 
F(1)=-5.2\% +3.5\% +0.2\% =-1.5\% $. This means that the 
reaction is only $1.5\% $ slower in the Sun than predicted by the 
Salpeter formula. Table 2 also compares our corrections with 
those predicted by Graboske, DeWitt, Grossman, \& Cooper (1973)
(GDGC), by Salpeter \& Van Horn (1969) (SVH), and 
by Dzitko, Turck-Chi\`eze, Delbourgo-Salvador, \& Lagrange (1995) (DTDL).

Our corrections are typically an order of magnitude 
smaller than the corrections calculated by 
GDGC (cf. columns two and three of Table~2). 
The intermediate screening prescription 
of GDGC (their Table 4, page 465) uses the intermediate screening formula
from DeWitt, Graboske, \& Cooper (1973) (their Eq. (70), page 455); 
the DeWitt et al. formula 
was obtained as an illustration assuming completely 
degenerate electrons, which is inappropriate for the solar interior. 
The GDGC intermediate screening 
prescription underestimates the enhancement of fusion reactions by a
factor,
$\exp(\delta \Lambda_{\rm GDGC} - \delta \Lambda_{\rm GB}) $, 
which  varies from about $7$\% for the 
important $^3{\rm He}(^4{\rm He},\gamma)^7{\rm Be}$ 
and $^7{\rm Be}({\rm p},\gamma)^8{\rm B}$ reactions  to about 
$16$\% for the $^{14}N({\rm p}, \gamma)^{15}O$ reaction (cf. Table~II
of Ricci et al. 1995).

The discrepancies between our results and those of Dzitko et
al. (1995) only become large when relatively  heavy nuclei are
involved.  In this
case, the electron density in the vicinity of the 
fusing nuclei is much larger than the value, 
$n_e(\infty)$, assumed by \hbox{Dzitko et al.}
Table~II shows, for example,
that when the electron density is calculated from the density matrix
equation the  value for $-\delta \Lambda$ is only 
$0.8$\% instead of the Dzitko et al. value of $6.3$\% 
(cf. columns two and five of Table~2).

For heavier nuclei like nitrogen, the large classical enhancement of
electron density near the nucleus competes with the smearing effect
due to quantum fuzziness, resulting in a net correction that is
smaller than for the lighter nuclei (see Figure~1).

\section{Summary and Conclusion}
\label{secconclusion}

We use the density matrix equation to determine from quantum
statistical mechanics the electron density in the near vicinity of the
fusing nuclei. Our treatment is the first to describe properly the
electron density in screening calculations that are appropriate 
for solar interior conditions. Previously, the 
lack of understanding of what to use for the
electron density near the fusing nuclei 
has been the principal cause for uncertainty in
estimating non-linear corrections to screening calculations 
(see, e.g., Ricci et al. 1995 and references therein).

The non-linear corrections that we calculate to the Salpeter weak screening
 formulae,
Eq.~(\ref{deff})--Eq.~(\ref{defnzeta}),  are,
for solar conditions, $\sim 1\% $ for all the important nuclear 
fusion reactions. 
The principal uncertainty in our calculations is caused by thermal
 fluctuations, which are not included in the present treatment.
  For the analogous case of
 electron capture, thermal fluctuations affect the average rate by 
$\leq 1$\% (Gruzinov and Bahcall 1997).  Since the non-linear effects
 calculated in the present paper are small and of the same order 
as the effects of fluctuations that occur in the electron capture
 problem, 
we recommend using the Salpeter weak screening formula for 
solar fusion rates. 

What difference do the present results make for the solar neutrino 
problem? This question is answered by Table 3 of 
Bahcall \& Pinsonneault (1992). Keeping all other input 
data constant, the weak screening approximation gives, relative to 
the Graboske et al. prescription,  
a 0.4 SNU larger result in the 
chlorine (Homestake) experiment, a 2 SNU increase in the gallium 
experiments, and a 7\% larger $^8$B neutrino flux 
(measured in the Super Kamiokande, Kamiokande, and SNO experiments). 
The Graboske et al. prescription 
was used previously by Bahcall
and Pinsonneault and in many other stellar evolution codes (cf. Ricci
et al. 1995).

An error in the screening enhancement is equivalent to an error in the
low energy cross section factor. 
Therefore, one can use the well known power law
dependences of the neutrino fluxes on cross section factors (Bahcall
1989) to
estimate the uncertainties introduced by inaccuracies  in the
screening calculations.
A 1\% uncertainty in the screening calculation
causes an  $\sim 1\% $ uncertainty in the 
predicted $^8$B neutrino flux and a  smaller uncertainty 
for other fluxes in the pp chain.
For the crucial $^8$B neutrino flux, the uncertainty in the
measurement of the low
energy cross section factor for the $^7Be({\rm p},\gamma)^8B$ reaction
causes a much larger uncertainty, $> 10$\% (see Bahcall and
Pinsonneault 1995).

The non-linear effects in ion and electron  screening 
that are evaluated in this paper cause differences in the
solar model  neutrino fluxes that are small compared 
to the order-of-unity 
differences between the rates measured in solar neutrino experiments
and the fluxes predicted by standard models (assuming nothing
happens to the neutrinos after they are created).

\acknowledgements

This work was supported by NSF PHY-9513835.  We are grateful to
S. Turck-Chi\`eze for valuable discussions that first directed
our attention to the problem of the large apparent
corrections implied by the Graboske et al. prescription and
stimulating comments on a draft of this manuscript.
We are grateful to M. Br\"uggen for a valuable discussion.

\appendix 
\section{Quantum Corrections to the ${3/2}T$ Kinetic Energy Per Particle Rule}
In classical statistical mechanics, the kinetic energy of particles, 
interacting or noninteracting, in an external potential or in free space, is 
${3\over 2}T$ per particle. In quantum statistical mechanics, the kinetic 
energy at a given temperature depends on the external potential. This is 
obvious in the low-temperature limit: the kinetic energy of the ground state 
is positive if the external potential is not zero (the zero-point 
oscillations). 

Thermal electrons in an electrostatic field of a nucleus have kinetic energy 
larger than ${3\over 2}T$. The effect depends on $Z$, and reduces the reaction 
rates (as compared to Salpeter's weak screening rates). The correction to 
kinetic energy can be calculated if the diagonal of the density matrix (e.g., 
Feynman 1990) $\rho (r,\beta)\equiv \rho (r,r,\beta)$ is known,
\begin{equation}
\delta K=n_e(2\pi \beta)^{3/2}\int d^3r\{-\partial _\beta \rho - ({3\over 
2}\beta ^{-1}+V)\rho \} ,
\end{equation}
that is the correction to the kinetic energy is the total energy minus the 
unperturbed kinetic energy, ${3\over 2}T$, minus the potential energy, $V$. In 
classical statistical mechanics 
\begin{equation}
\rho =(2\pi \beta)^{-3/2}e^{-\beta V},
\end{equation}
and Eq. (A1) gives $\delta K =0$. 

We calculated the density matrix of electrons analytically and used Eq. (A1) 
to calculate the kinetic energy correction for $V=-{Z\over r} e^{-r/R_D}$ 
assuming $\beta \ll 1$, $Z\sim $ few, $R_D>\beta ^{1/2}$. These conditions are 
satisfied in the solar interior where $\beta \approx 0.02$ and  $R_D\approx 
0.5$. Two different approaches were used at distances from the nucleus greater 
than the de Broglie wavelength,  $\beta ^{1/2}$, and at distances smaller than the 
Debye radius $R_D$. These two approaches are explained below.

\subsection{$r\gg \beta ^{1/2}$: High-Temperature Expansion}

Thermal electrons have ``a characteristic size'' $\sim \beta ^{1/2}$. If the 
potential energy does not change by much over this distance (which in our case 
is true for $r> \beta ^{1/2}$), the density matrix is approximately given by 
Eq. (A2) with small corrections. The corrections are due to the fact that a 
fuzzy thermal electron samples potential not only at a given point but in the 
$\beta ^{1/2}$-vicinity of the given point.

Let $(x,y,z)$ be a small deviation of coordinates from $(r,0,0)$. Potential 
energy is, up to the second order,
\begin{equation}
\delta V= V'x+{1\over 2}V''x^2+{1\over 2}{V'\over r}y^2+{1\over 2}{V'\over 
r}z^2,
\end{equation}
where prime denotes the $r$-derivative, and we assume that $V$ is spherically 
symmetrical. The path integral giving the density matrix (e.g., Feynman 1990, 
chapter 3) is Gaussian and can be calculated. In fact, the answer can be 
constructed without the actual calculation from the known density matrix of 
the linear harmonic oscillator (e.g., Feynman 1990, chapter 2). It reads 
\begin{equation}
\rho =(2\pi \beta)^{-3/2}e^{-\beta V}\{1+{1\over 24}\beta ^3 V'^2-{1\over 
12}\beta ^2(V''+{2\over r}V')\} .
\end{equation}
The kinetic energy correction is given by Eq. (A1)
\begin{equation}
\delta K=n_e\int 4\pi r^2dre^{-\beta V}\{-{1\over 8}\beta ^2V'^2+{1\over 
6}\beta (V''+{2\over r}V')\} .
\end{equation}
In our case Eq.(A4) is valid only at $r\gtrsim  \beta ^{1/2}$, but if 
potential energy $V$ were smooth at all $r$, we could have integrated the last 
term by parts 
\begin{equation}
\delta K={1\over 24} n_e\beta ^2\int 4\pi r^2dre^{-\beta V}V'^2,
\end{equation}
showing that kinetic energy correction is positive in the high-temperature 
limit. In our calculation we used Eq. (A4) at $r>r_0$, and results from the 
next section were used at $r<r_0$. The final answer does not depend on the 
choice of $r_0$ as long as $R_D\gtrsim r_0\gtrsim \beta ^{1/2}$.

\subsection{$r\ll R_D$: Hydrogenic Density Matrix}

At distances from the screened nucleus $r\ll R_D$, the potential energy is
\begin{equation}
V=-{Z\over r} \exp (-{r\over R_D})\approx -{Z\over r}+{Z\over R_D}.
\end{equation}
The only effect of the constant correction ${Z\over R_D}$ is to lower electron 
density by the Boltzmann factor $e^{-\beta Z/R_D}$. The density matrix in the 
Coulomb potential can be obtained from hydrogenic eigenstates. 

The kinetic energy correction is 
\begin{equation}
\delta K=n_ee^{-{\beta Z\over R_D}}(2\pi \beta)^{3/2}\int d^3r\{-\partial 
_\beta \rho - ({3\over 2}\beta ^{-1}+V)\rho \} .
\end{equation}
The density matrix is 
\begin{equation}
\rho (r,\beta)=\sum_{l=0}^{\infty } {2l+1\over 4\pi }\{ \sum_{n=1}^{\infty } 
|R_{nl}(r)|^2e^{\beta \over 2n^2}+\int_0^{\infty }{dk\over 2\pi 
}|R_{kl}(r)|^2e^{-{\beta k^2\over 2}}\} .
\end{equation}
Here the bound states of hydrogen are (e.g. Landau \& Lifshitz 1977) 
\begin{equation}
R_{nl}(r)={2\over n^{l+2}(2l+1)!}\{ {(n+l)!\over (n-l-1)!}\} 
^{1/2}(2r)^le^{-r/n}F(-n+l+1,2l+2,2r/n),
\end{equation}
where $F$ is the confluent hypergeometric function. The continuum states are 
\begin{equation}
R_{kl}(r)=2ke^{\pi /2k}|\Gamma (l+1-{i\over k})|(2kr)^le^{-ikr}F({i\over 
k}+l+1,2l+2,2ikr),
\end{equation}
and for $Z\ne 1$ we scale $r\rightarrow Zr$, $\beta \rightarrow Z^2\beta$. 

We used these formulae to calculate the kinetic energy shift at small $r$. 
Results of this subsection match the high-temperature results if $\beta 
^{1/2}<r<R_D$. We repeated the calculation at smaller $\beta $ to obtain the 
free energy shift due to the quantum correction to kinetic energy of electrons,
\begin{equation}
\beta \delta F=\int_0^{\beta }d\beta '\delta K (\beta ').
\end{equation}
Results are shown in Table 1.

\newpage

\begin{deluxetable}{lcccccc}
\footnotesize
\tablecaption{Electrostatic, Kinetic and Free Energy Corrections} 

\tablehead{Z&1&2&4&5&7&8}
\startdata
$\beta \delta U, ~~\% $   &0.34&1.6&6.4&9.2&11.2&7.6\nl
$\beta \delta F_U, \% $   &0.1&0.6&2.7&3.9&5.7&5.2\nl
$\beta \delta K, ~~\% $   &0.22&0.57&1.9&3.2&8.1&12.6\nl
$\beta \delta F_K, \% $   &0.1&0.3&0.8&1.3&2.9&4.4\nl
$\beta \delta F, ~~\% $   &0.2&0.9&3.5&5.2&8.6&9.6\nl

\enddata

\tablecomments{The symbols represent: nuclear charge $Z$, corrections to: (i) 
the electrostatic energy normalized to temperature $\beta \delta U$, (ii) the 
free energy due to increased electrostatic energy $\beta \delta F_U$, (iii) 
the kinetic energy of electrons $\beta \delta K$, (iv) the free energy due to 
increased kinetic energy $\beta \delta F_K$, (v) the total free energy $\beta 
\delta F$. The plasma parameters are taken from the solar model of Bahcall \& 
Pinsonneault 1995 at the representative point $R/R_{\odot }=0.06$.}

\end{deluxetable}

\begin{deluxetable}{lccccc}
\footnotesize
\tablecaption{Reaction Rate Corrections} 

\tablehead{Reaction&GB&GDGC&SVH&DTDL}
\startdata
p+p   &0.5&0.0&0.5&0.2\nl
$^3$He+$^4$He &1.7&8.2&2.4&1.8\nl
p+$^7$Be  &1.5&8.5&2.6&2.3\nl
p+$^{14}$N   &0.8&15.2&6.3&6.3\nl

\enddata

\tablecomments{Corrections  to weakly screened reaction rates.
Nuclear fusion reactions in the sun are enhanced by a factor 
$\exp (\Lambda + \delta \Lambda)$, where $\Lambda$ is given by 
Salpeter's expression, which is Eq.~\ref{deflambda}.  The table shows the
corrections, $-\delta \Lambda $ in percent, 
calculated in this paper (GB), by Graboske, DeWitt, Grossman, \& Cooper
(1973) (GDGC), by Salpeter \& Van Horn (1969) (SVH), and by Dzitko,
Turck-Chi\`eze, Delbourgo-Salvador, \& Lagrange (1995) (DTDL). 
The corrections refer to a representative point $R/R_{\odot} = 0.06$.}

\end{deluxetable}

\begin{figure}[htb]
\psfig{figure=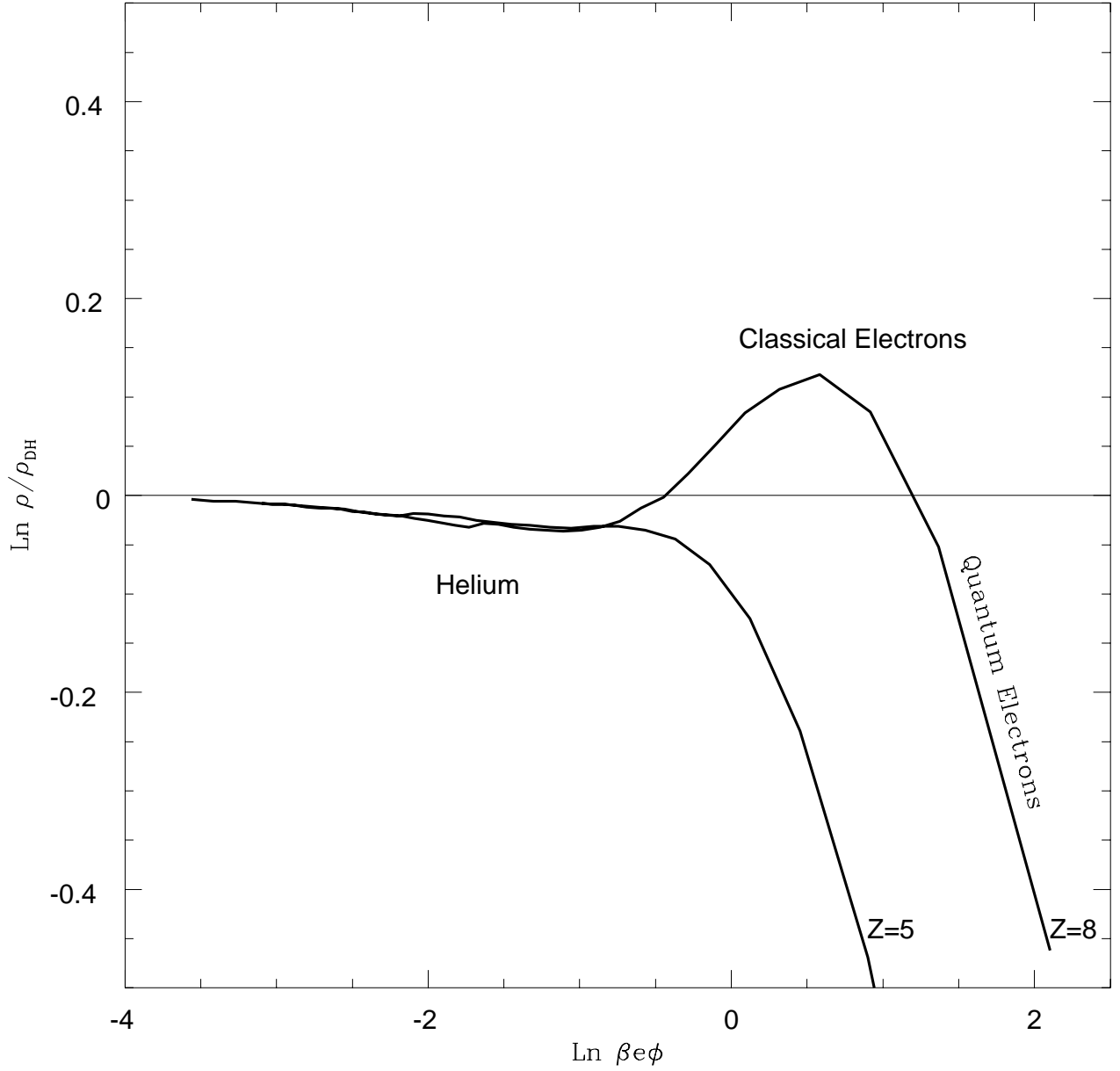,width=7in}
\caption{Screening of two test charges. 
The induced charge density, $\rho$, normalized to the Debye-H\"uckel
charge density, $\rho_{\rm DH} = \phi/(4\pi R_D^2) $, is shown as a
function the electrostatic potential, $\phi$. 
For small $\phi$ (large distances from the screened nucleus), $\rho$
is given by the classical Boltzmann formula (see the right hand side
of Eq.~\ref{eqnPB} ).  In this region, $\rho$ is smaller than
$\rho_{\rm DH}$ due to the presence of helium ions. At large $\phi$
(close to the screened nucleus), $\rho$ is much smaller than
$\rho_{\rm DH}$  due to the quantum fuzziness embodied in the density
matrix equation (Eq.~\ref{densitymatrixeq}).  At intermediate
distances, $\rho$ can be large  than
$\rho_{\rm DH}$ (because $\exp\phi > \phi$ ).
For Z = 8 the plasma response is larger than linear at intermediate
distances. For Z = 5 the plasma response is always smaller than linear.}

\end{figure}

\end{document}